# Neural network prediction of load from the morphology of trabecular bone

Amir Abbas Zadpoor[1,2,*], Gianni Campoli[2,*], Harrie Weinans[*,†]

[*]*Department of Biomechanical Engineering, Faculty of Mechanical, Maritime, and Materials Engineering, Delft University of Technology (TU Delft), Mekelweg 2, Delft 2628 CD, The Netherlands*
[†]*Department of Orthopaedics, Erasmus University Medical Center, Rotterdam, The Netherlands*

**ABSTRACT**

Bone adaptation models are often solved in the forward direction, meaning that the response of bone to a given set of loads is determined by running a bone tissue adaptation model. The model is generally solved using a numerical technique such as the finite element model. Conversely, one may be interested in the loads that have resulted in a given state of bone. This is the inverse of the former problem. Even though the inverse problem has several applications, it has not received as much attention as the forward problem, partly because solving the inverse problem is more difficult. A nonlinear system identification technique is needed for solving the inverse problem. In this study, we use artificial neural networks for prediction of tissue adaptation loads from a given density distribution of trabecular bone. It is shown that the proposed method can successfully identify the loading parameters from the density distribution of the tissue. Two important challenges for all load prediction algorithms are the non-uniqueness of the solution of the inverse problem and the inaccuracies in the measurement of the morphology of the tissue. Both challenges are studied, and it is shown that the load prediction technique proposed in this paper can overcome both.

## 1. INTRODUCTION

It is well established that the form and function of bones are linked such that the mechanical properties and microstructure of bones are dependent on their loading history. It is of immense practical importance to be able to predict the response of bones to a given set of loading. Examples of the areas where such predictive capability is of great value are the design and post-operative analysis of orthopaedics implants [1-4], study of fracture healing [5-8], and prediction of the bone adaptation caused by disuse or prolonged exposure to microgravity [9-12]. In such cases, the loading of bones is more or less known and one needs to determine the response of a bone to the specified loading. This is an example of the so-called 'forward modeling' and has been extensively studied in the literature. Conversely, one may be interested in determining the loading experienced by a bone given the current state of the tissue. The actual state of bone is (at least partly) determined by the loads the bone experiences. An inverse problem should be solved: what are the inputs that have resulted in this given output? The solution of the inverse problem provides us with a non-invasive method for estimating the dominant loading patterns of joints [13] and has several practical applications. For example, it can be used for noninvasive estimation of the musculoskeletal loads and the characteristic of the daily activities that have resulted in a measured density distribution. Bona proposed a contact algorithm for density-based load estimation and used the method to distinguish between different modes of locomotion of various animals [14]. As another example, Fischer *et al.* showed that density-based load estimation predicts altered femoral load directions for coxa vara and coxa valga [15].

---
[1] Corresponding author, email: a.a.zadpoor@tudelft.nl, tel: +31-15-2786794, fax: +31-15-2784717.
[2] Both authors have equally contributed to this manuscript and should therefore be considered as joint first authors.

Even though the inverse tissue adaptation problem is fundamentally important and has many applications, it has not been extensively studied so far, presumably due to the difficulties that are associated with the solution of the nonlinear inverse problem. The most important works in this area include a series of studies by Fischer and co-workers [13, 15-20]. In the vast majority of the previous works, an iterative load prediction technique specific to the problem under study was used for prediction of the involved load cases.

In this paper, we use artificial neural networks (ANNs) for prediction of loads from the spatial distribution of density. ANNs have been recently used in conjunction with finite element modeling to study different aspects of bone tissue adaptation including multiscale simulation of bone remodeling [21] and damage accumulation in trabecular bone [22]. In the current study, we used ANNs to predict the loads experienced by trabecular bone. The idea of using ANNs and finite element modeling for solving inverse problems has been previously used in other engineering fields such as metal forming [23]. However, ANNs have not been previously used in conjunction with bone tissue adaptation theories for prediction of the load experienced by bone. Therefore, the most important contribution of the current study is proposing and testing a new approach for predicting the load from the density distribution of bone. No efficient and universally applicable technique is currently available for solving this practically important problem.

One needs to solve an inverse problem in order to predict the load experienced by bone from its measured density distribution. As the mapping from the space of density distribution to the space of loading parameters is a nonlinear mapping, only nonlinear techniques can be used for solving this inverse problem. ANNs are among the nonlinear system identification methods that can be used for this purpose. An artificial neural network is, in principle, a nonlinear mapping from the space of inputs (e.g. density distribution) to the space of outputs (e.g. loading parameters). ANNs have several advantages compared to other nonlinear system identification techniques. First, ANNs are very general. It is proven that ANNs can accurately represent *any* sufficiently smooth nonlinear mapping [24]. ANNs can therefore be used for prediction of any type of load in any form of tissue adaptation process including both long (whole) bones and trabecular bones and (at least theoretically) any other tissue. Second, the accuracy of the solution is independent from the number of inputs [24]. This is an important point, because prediction of the load from density distribution may require introducing an accurate description of the density distribution to the ANN. A large number of inputs are normally required for accurately describing the density distribution. Third, ANNs are particularly useful in the cases where solving the forward finite element model is time consuming [25, 26]. Since solving the forward tissue adaptation problem can be quite time consuming particularly when the number of elements is too large or the loading pattern is very dynamic. ANNs can help us reduce the time required for predicting the load, because generation of the training dataset that involves solving the forward finite element problem needs to be done only once. Afterwards, the trained ANN can be used for load prediction without any need for solving the forward problem.

The problem that is studied here is designed to be close to the worst-case scenario. Two important aspects in the estimation of the load from the measured density distribution are the non-uniqueness of the solution of the inverse problem and the inaccuracies that may exist in the measurement of the density distribution. It has been long known that very different sets of loads can result in very similar density distributions [27]. The non-uniqueness of the solutions of the inverse problem is a challenge for any load prediction algorithm, because several solutions exist for the same inverse problem. The convergence of the solution may therefore be compromised. The second challenge is the inaccuracies that may exist in the actual measurements of bone density distribution. The load prediction algorithm should be robust enough to be able to provide reasonable predictions of load even when the density distribution measurements are not perfectly accurate. Both challenges are addressed for the proposed load prediction algorithm.

## 2. METHODOLOGY

Two sets of tools are needed for prediction of the loads applied to a particular tissue. First, one needs a model of the tissue adaptation process implemented in a numerical modeling platform such as a finite element computer program. Second, a technique is needed for prediction of loads based on the governing equations of the tissue adaptation model. The present section of the paper is divided into three sub-sections. The first sub-section presents the tissue adaptation model and the last two sub-sections present the load prediction algorithm.

### 2.1. Tissue adaptation model

The tissue adaptation model used in this study is a version of the strain energy density model proposed and used by Huiskes and coworkers [28-32]. The model takes a physiological approach and assumes that a network of sensors inside the bone tissue (e.g. osteocytes) generate the remodeling stimulus signal. Based on this assumption, a complex exchange of mostly chemical information leads to change in the morphology of the trabecular bone tissue that embodies the sensors. Bone tissue is considered as an isotropic material. The Young's modulus of the tissue is considered to be a function of its density, while the Poisson ratio is fixed at 0.3.

The bony area is assumed to have $N$ sensor cells, uniformly distributed over its volume. Every element of the finite element model is assumed to contain one sensor. Any sensor $i$ measures the stimulus signal, $S_i$, which in this specific case is the strain energy per unit of mass, calculated at the location of the sensor as:

$$S_i = \frac{U_i}{\rho_i} \qquad (1)$$

where $\rho_i$ is the density at the location of the sensor. The strain energy density, $U_i$, is calculated from the stresses and strains sensed by the sensor:

$$U_i = \frac{1}{2}\boldsymbol{\sigma}_i : \boldsymbol{\varepsilon}_i \qquad (2)$$

The density $\rho_i$ at location $x$ is regulated by the stimulus signal, $\varphi(x,t)$, to which all sensor cells contribute. The contribution of every sensor depends on its distance from the location $x$. For every time increment, $\Delta t_i$, the difference between the strain energy density and a reference value, $k$, is calculated:

$$\varphi(x,t) = \sum_{i=1}^{N} f_i(x)(S_i - k) \qquad (3)$$

where $N$ is the number of elements in the FEM model. If the difference is zero, the process stops. Otherwise it adds or removes material according to the following density modification rule:

$$\frac{d\rho(x,t)}{dt} = \tau\varphi(x,t) \qquad (4)$$

where $\tau$ is a time constant that regulates the rate of the process. The algorithm stops adding density when it reaches a given maximum density, $\rho_{max}$, and stops removing density when it reaches the minimum density, $\rho_{min}$. The spatial influence function, $f_i(x)$, determines how the contribution of the sensors changes with distance from the location, $x$. Following Mullender *et al.* [33], it is assumed that the contribution of sensors decays exponentially with distance from the location, $x$:

$$f_i(x) = e^{-d(x)/D} \qquad (5)$$

where $d(x)$ is the distance between the sensor $i$ and the location $x$. The rate of spatial decay is regulated by the parameter $D$. The elastic modulus at location $x$ is calculated from the density:

$$E(x,t) = C\rho(x,t)^\gamma \qquad (6)$$

where $C$ and $\gamma$ are material constants.

The bone adaptation model is solved using an implicit finite element solver (ABAQUS standard). The finite element (FE) model includes the relation between the density and Young's modulus. The model is solved in a quasi-static process: the loads are applied and the resulting stress and strains are calculated. The stress and strain fields are then used by an algorithm that calculates the remodeling stimulus signal and updates the density and mechanical properties of the tissue according to the stimulus signal.

The tissue adaptation problems considered in this study (Figure 1) are similar to the ones studied in [28, 32, 33]. The loading, boundary conditions, and sizes are presented in Figure 1. Figure 1a-b present the two different types of loading that are used here. The first type of loading (Figure 1a) includes a linearly varying line load similar to the one used in [32]. The second type of loading is a combination of four constant line loads applied on the edges of the geometry (Figure 1b).

The relationship used for calculation of the remodeling stimulus signal is slightly modified in this study. According to Equation (3), the intensity of the stimulus signal increases as the number of sensors (elements) increases. However, this is not necessarily an accurate assumption. For a sufficiently accurate FE representation of the tissue adaptation model, the remodeled shape should be independent from the number of elements. Moreover, the remodeled shaped should not be oversensitive to the parameter $D$. The problem illustrated in Figure 1a was simulated using the remodeling stimulus signal described by Equation (3). The parameters that were used for this simulation are presented in Table 1. The results of this simulation (Figure 2a) show that if the stimulus signal is calculated according to Equation (3), the outcome of the tissue adaptation model is very much dependent on the choice of the parameter $D$. For sufficiently large $D$ values (e.g. $D = 0.125$), the stimulus signal is so much diffused that there is no difference between the different locations of the simulated geometry. Furthermore, the simulation results are mesh-dependent, because a finer mesh increases the number of elements and, thus, the intensity of the stimulus signal. In order to avoid these problems, the formulation of the stimulus signal was modified as follows:

$$\varphi(x,t) = \frac{\sum_{i=1}^{N} f_i(x)(S_i - k)}{\sum_{i=1}^{N} f_i(x)} \qquad (7)$$

where $N$ is the number of elements in the FEM model. In this way, the stimulus signal is normalized with respect to the sum of the values of the spatial influence function. This modification solves the above-mentioned problems about the over-sensitivity of the solution to the parameter $D$ and mesh-dependency of the solution. The simulation presented in Figure 2a was repeated this time using the remodeling stimulus signal described by Equation (7). The results of the new simulation (Figure 2b) showed that the solution of the modified tissue adaptation model is much less dependent on the choice of $D$. Moreover, it was observed that the simulation results are mesh-independent.

**2.2. Load prediction algorithm**

In general, there is a nonlinear mapping from the space of the loads applied to the bone to the space of the response of the tissue. The response of the bone to the applied load is represented in our model as the spatial distribution of the density (porosity) of the bone. This first mapping is called 'forward mapping', because this mapping goes forwards in time and predicts the spatial density distribution that results from a given set of loading parameters. The inverse of the forward mapping goes backward in time. Given a certain spatial density distribution, the so-called 'backward mapping' determines the set of loading parameters that has resulted in the given spatial distribution of the density, thereby mapping the space of density distribution to the space of loading parameters. The forward mapping can be easily constructed by implementing the tissue adaptation algorithm in the FEM model. For any set of loading parameters, the FEM model calculates the resulting spatial distribution of the density based on the solution of the tissue adaptation equations described in the previous sub-section. However, there is no straightforward way of constructing the backward mapping. That is because the governing differential equations of the backward mapping are not known. One therefore has to resort to nonlinear system identification techniques for identification of the backward mapping. As far as the system identification techniques are concerned, there are two difficulties associated with identification of the backward mapping. First, the morphology of the forward and, consequently, backward mapping is complex. Second, an accurate description of the spatial distribution of the density requires use of a large number of input variables. Moreover, the number of output variables is quite large (4 loading parameters in the case of the problem considered here). In this study, Artificial Neural Networks (ANNs) were selected to establish the backward mappings. This choice was motivated by the special properties of ANNs. It is mathematically proven that a feedforward ANN with at least one hidden layer, $n$ hidden neurons, and sigmoid activation functions can approximate any continuous function with an integrated squared error of the order $O(\frac{1}{n})$ regardless of the dimension of the input space [34]. The point that approximation error is independent from the dimension of the input space relieves the difficulty with the large number of data points that are needed for accurate representation of the spatial density distribution. Furthermore, every continuous function can be approximated to an arbitrary degree of accuracy, meaning that the ANNs are capable of successfully representing the complex topology of the backward mapping.

*2.2.1. Artificial neural networks*

ANNs are mathematical models that are inspired by the architecture and/or function of biological neural networks [35]. They are composed of the so-called artificial neurons that are the building blocks of ANNs. The way these neurons are connected to each other and the way they are trained determine the structure and function of any particular category of ANNs. Various architectures are proposed for ANNs including feedforward, radial basis function (RBF), counterpropagation, and learning vector quantization (LVQ) networks. For a description of the different architectures and training algorithms of ANNs see the recent review paper by Wilamowski [36]. Feedforward ANNs and a backpropagation training algorithm were selected for the current study because of the reasons mentioned in the previous paragraph.

Figure 3a presents a schematic drawing of a typical feedforward ANN. A typical feedforward ANN has a certain number of inputs, $n_{in}$, and a certain number of outputs, $n_{out}$. ANNs map the space of inputs to the space of outputs. Among the layers of feedforward ANNs one layer is the input layer and receives the input to the network and another is called the output layer and returns the output of the network. The layers that are between the input and output layers are called hidden layers. Each hidden layer contains a number of so-called hidden neurons. The numbers of neurons in the input and outputs layers are respectively equal to the number of inputs and outputs of the ANN. However, hidden layers may have as many hidden neurons as needed. The processing job is carried out in the hidden layers. Input and output layers only connect the ANN to the outside world. Various layers and neurons are strongly interconnected (Figure 3a).

A representative neuron is depicted in Figure 3b. Every such neuron has a so-called activation function, $f$, such as the tang-sigmoid function. The neuron assigns weights, $w_i$, to the signals, $p_i$, that come to it through its incoming connections. The weighted signals are then summed up and biased (summed with $b$) and the result is introduced to the activation function. The scalar outcome of the function is sent out of the neuron through the outgoing connections of the neuron.

Application of ANNs requires that the parameters of neurons (weights, $w_i$, and biases, $b$) be tuned such that the mapping carried out by the ANN is the same as or very close to the desired mapping. The process of tuning the parameters of an ANN is called training. A training dataset composed of a number of inputs and their corresponding (target) outputs is needed for this purpose. The ANN is trained using a training algorithm such that, for the given set of inputs, the outputs of the ANN are as close to the target outputs as possible. In this study, the training of the ANNs was in most cases carried out using 90% of the training data points and by applying the quasi-Newtonian back-propagation method of Levenberg-Marquardt. The remaining 10% of the training data points were divided into two 5% subsets that were used for the validation and testing of the trained ANN. The split into training, validation, and test datasets was done on a random basis. The validation dataset is not used in the training of the ANN but is used to evaluate the improvement of the generalization capability of the ANN while it is being trained. The training is continued as long as the generalization capability of the ANN is improving and is stopped when generalization is not improving anymore, meaning that the ANN is being over-fitted to the training dataset.

The test dataset is not used in the training process at all. It can be therefore used to assess how the ANN performs for data points other than the ones used for its training or validation. Regression parameter, $R$, is defined as the Pearson correlation coefficient between two vectors: one containing the actual values of the four loading parameters and the other containing the identified values of the loading parameters.

In summary, the ANN that is used in this study receives the density distribution as input and returns the loading parameters, $F1-F4$, as output. The ANN should be trained using a training dataset. The training dataset consists of a series of runs of the forward model for a number of randomly chosen loading parameters within the ranges that are specified for the loading parameters. The simulation of the forward model was carried out for 200 iterations after which there were no significant changes in the predicted density distribution. For this training dataset, both the density distribution and the loading parameters are known. The ANN is trained such that for any given density distribution in the training dataset, it returns the values that are as close to the corresponding loading parameters as possible. The number of inputs introduced to the ANN is equal to the number of elements used in the forward FEM model. The input data is not normalized and is introduced to the ANN as a vector. Only one hidden layer was used in the design of all ANNs used in the current study.

## 2.3. Algorithms for dealing with non-unique solutions of the inverse problem

As already mentioned, the solution of the inverse problem considered here is not unique. The non-uniqueness of the solution is an important challenge for all load prediction methods including the method proposed in the current study. Two algorithms are proposed here for overcoming the challenges presented by the non-uniqueness of the solution. The main idea is to divide the load prediction task between two ANNs instead of using one ANN. While the first ANN is used for prediction of unique loads, the second ANN is used for prediction of non-unique loads. Each of the two proposed algorithms works with one of these two ANNs.

The first algorithm is a simple algorithm for identifying and deleting the non-unique solutions from the training dataset. The first step is to calculate the similarity between all density distributions. The norm of the vector connecting two given density distributions $i$ and $j$ is used as an indicator of their similarity:

$$l = \|\rho_i - \rho_j\| \qquad (8)$$

where $\boldsymbol{\rho} = \begin{bmatrix} \rho_1 & \rho_2 & \rho_3 & \ldots & \rho_n \end{bmatrix}$, $n$ is the number of the elements of the finite element model, and $i$ and $j$ are the identification numbers of the two density distributions that are being compared.

The second step is to delete all the density distributions whose distance from at least one other density distribution is less than a chosen threshold, $l_{th}$. The remaining density distributions are then used for training of the ANN. It is important to note that because all the non-unique density distributions are deleted from the training dataset, the ANN is only capable of identifying the unique loading parameters.

Non-unique loading parameters cannot be identified using the first ANN. A second algorithm is proposed for training of a second ANN (ANNII) that is used for prediction of non-unique loading parameters. The algorithm used for the training of the second network includes the following steps:

1. Calculate the distance between the density distribution $i$ and all other density distributions in the training dataset. Calculate the distance between the loading parameters associated with the density distribution $i$ and the loading parameters associated with all other density distributions. The distance between two sets of loading parameters is defined as:

$$L = \|\boldsymbol{F}_i - \boldsymbol{F}_j\| \qquad (9)$$

   where $\boldsymbol{F} = \begin{bmatrix} F1 & F2 & F3 & F4 \end{bmatrix}$.

2. For every density distribution $i$, find all the density distributions whose distance from density distribution $i$ is less than $l_{th}$ and the distance of their corresponding loading parameters from the loading parameters associated with density distribution $i$ is less than $L_{th}$. Put all such training samples, $j$, together with the training sample $i$ in a set $\mathbb{N}_i$. The use of a second distance threshold, $L_{th}$, guarantees that only genuinely non-unique solutions are captured not the density distributions that have simply resulted from loading parameters that are too close to each other.

3. Find sets, $\mathbb{N}_i$, that share members and replace them with their union. In this way, the non-unique solutions are classified into many classes. The density distributions of the members of every class are similar to each other.

4. Select one of the members of every class to represent that class in the training of the second ANN.

5. Train the ANN using the training dataset that consists of one representative member from every class.

The resulting ANN (ANNII) can be used for prediction of non-unique loading parameters.

## 3. RESULTS AND DISCUSSIONS

The results and their related discussions are presented in four subsections. The first subsection (Section 3.1.) presents general results. In this subsection, the non-uniqueness of the solution and the inaccuracy of the density measurements are not considered. The second and third subsections (Sections 3.2. and 3.3) respectively consider the non-uniqueness of the solution and the existence of noise in the measured density distributions. The last subsection discusses the implications of the study for biomechanical research.

### 3.1. Prediction of unique loading parameters from noise-free density distributions

For the results presented in this subsection, the ranges of the loading parameters ($F1 : 2.5 - 4$, $F2 : 1 - 2.5$, $F3 : 4 - 5.5$, and $F4 : 5.5 - 7$) were chosen such that the solution of the inverse problem was unique. Preliminary studies showed that application of even a coarse mesh ($5 \times 5$ elements) is adequate for obtaining an extremely accurate estimation of the loading parameters using an ANN. A $5 \times 5$ elements mesh was therefore used throughout this study. The results of the preliminary study also showed that 20 hidden neurons are sufficient for accurate estimation of the loads. The quality of the ANN was not oversensitive to the number of hidden neurons. Figures 4 and 5 respectively present the results of the training and testing of the ANN when 100 or 1000 training data samples were used. Even for a training dataset composed of only 100 samples (Figure 4), the load estimation error decreases by 3 orders of magnitude within less than 40 training iterations. When the number of training samples increases to 1000, the load estimation error decreases by 5 orders of magnitude within less than 100 training iterations. When the size of the training dataset is not very large (Figure 4), there is a slight difference between the performance of the ANN for estimation of the loads that were used in its training (Figure 4a), and the ones that were not used in the training of the network (Figure 4b-c). The improvement of the performance while training the network (Figure 4d) is also notably different between the training dataset on the one hand and the validation and test datasets on the other hand. However, when the size of the training dataset is large enough (Figure 5), there is not any notable difference between the accuracy of the ANN in estimation of the loads that were used in its training (Figure 5a) and the ones that were not used in the training and were only used for the validation and testing of the network (Figure 5b-c). The improvement of the performance while training the network is also very similar between the different parts of the training dataset (Figure 5d). The distribution of the identification error for the test dataset was close to a normal distribution (Figure 5e). Statistical analysis of those identification errors (Table 2) reveled that the mean identification error is very small (order of magnitude ≤ -3). The ratio of the mean absolute identification error was always less than 0.2% of the size of the identification interval (= 2.5–1 = 4–2.5 = 5.5-4 = 7-5.5 = 1.5).

The computational time required for completing one forward simulation is around 7 seconds on one single thread of an Intel® Core i7® CPU. Therefore, the time needed for generation of the training dataset is around 700 (Figure 4) or 7000 (Figure 5) seconds if only a single core is used for the simulations. An important point about the generation of training datasets is that the whole process is very easy to parallelize, because the generation of the training dataset consists of independently running the forward tissue adaptation model for a number of loading parameters. Therefore, one can run several simulations at the same time on the different threads/cores of the same CPU. In the case of our Core i7 CPU, the number of available threads was 8, meaning that the time required for completing the generation of the training dataset was 88(=700/8) and 880 (=7000/8) seconds. In summary, when the solution of the inverse problem is unique and the noise is nonexistent, the ANN performs extremely well in estimation of the loads even for a small training dataset of 100 samples. Moreover, very little computational resources are required.

**3.2. Non-uniqueness of the solution of the inverse problem**
In this subsection, the effects of the non-uniqueness of the solution of the inverse problem on the performance of the ANN in estimation of the loading parameters are studied. The range of loading parameters ($F1:2-5$, $F2:2-5$, $F3:2-5$, and $F4:2-5$) were modified to make non-unique solutions possible. Moreover, the sizes of the intervals within which the loading parameters vary were twice as large (3 instead of 1.5). The training dataset was generated by running the tissue adaptation model for a number of random combinations of parameters within the specified ranges. In a way similar to the previous subsection, 1000 training samples were used for the training of the ANN (Figure 6). The number of hidden neurons was not constant but was optimized using an interval-scanning algorithm that was aimed at minimizing the prediction error for the test dataset. One can see that there is a significant loss of prediction accuracy, if the training dataset includes non-unique solutions. That is because the training of the ANN is disrupted by the non-uniqueness of the solution. When the solutions are unique, the parameters of the ANN are adjusted during the training process such that for any given density distribution, the outputs of the ANN are as close to the loading parameters as possible. When several sets of loading parameters result in the same density distribution, the training algorithm will adjust the parameters of the ANN such that the sum of the prediction errors of all non-unique solutions is minimized. The outputs of the ANN are therefore somewhere between the different sets of loading parameters that result in the same density distribution. Some trials showed that the identified loading parameters for a given density distribution are close to the average of the non-unique loading parameters that result in that density distribution. In order to solve this problem, the algorithm presented in section 2.3 for deleting the non-unique solutions of the inverse problem was used. A question raised here is that 'how large should the distance threshold be?' The chosen threshold should be neither too large nor too small. An excessively large distance threshold will result in too many samples being deleted. The number of the training samples that are used for the training of the network would be therefore limited. A very small distance threshold will leave a large number of similar density distributions in the training dataset, meaning that the non-uniqueness problem would persist. In order to understand the effects of the chosen threshold on the number of deleted training data points, a large study with 65000 data points was conducted. Different distance thresholds varying between $10^{-6}$ and 0.09 were used (Figure 7). It was observed that the plot of the percentage of the deleted samples vs. distance threshold is a bi-linear graph. As the distance threshold increases, the number of deleted samples increases with a slow linear trend up to an elbow point (≈0.03 in the case of this example). After the elbow point, the slope of the linear trend drastically increases. The highly different slopes of the two linear trends imply that different types of training data points are deleted before and after the elbow point. While for the distance thresholds lower than the elbow point mostly non-unique solutions are deleted, genuinely unique solutions seem to be deleted after the elbow point. It is therefore suggested that the distance threshold should be chosen close to the elbow point. A distance threshold slightly larger than the elbow point guarantees that all non-unique solutions are deleted and not many training data points are lost. The performance of the ANNs that were trained using unique data points (Figure 8) were found to be much better than the ANNs that were trained using non-unique training data points (Figure 6). The performance of the ANN improved once the distance threshold increased (compare Figure 8a-c with Figure 8d-f).

As non-unique solutions were deleted from the training dataset of the last ANN, this ANN is not capable of identifying the loading parameters when the presented density distribution are resulted from non-unique sets of loading parameters. In order to be able to identify the loading parameters even when the solution of the inverse problem is non-unique, the second algorithm proposed in Section 2.3 was applied for training of a second ANN (ANNII). It was observed that ANNII is performing well in prediction of non-unique loading parameters (Figure 9). It can therefore be concluded that the proposed algorithm can provide a good basis for training of an ANN specialized in prediction of non-unique loading parameters. The important point about the proposed algorithm is that once the loading parameters are identified using ANNII, one can use the classes (sets) created in the pre-training phase as a look-up table for finding the other solutions of the inverse problem. In that case, the identified values are compared with the members of the classes. If a member of a class is found to be sufficiently close to the identified set of loading parameters, the other members of the same class can be regarded as the other solutions of the inverse problem. By running the forward model, one can check whether or not the other members of the class result in sufficiently close density distributions.

### 3.3. Robustness of the load prediction algorithm

In reality, the measurements of density distribution are not noise-free. It is therefore important that the load prediction algorithm is robust enough to handle the noise that will exist in actual density measurements. In order to test the robustness of the ANN when identifying the loading parameters from noisy data, the trained ANN was tested using noisy density data. Noise was introduced to the density distribution as follows:

$$\rho'(x) = \rho(x) + \mathcal{N}(0, \rho(x)/\lambda) \qquad (10)$$

where $\mathcal{N}(\mu, \sigma)$ stands for a Gaussian distribution with the mean $\mu$ and standard deviation $\sigma$. The signal to noise ratio is represented by $\lambda$.

An ANN similar to the one discussed in Section 3.1 was trained using a noise-free dataset. The ANN was then tested using a slightly noisy dataset (Figure 10, $\lambda = 100$). It was found that the ANN is extremely sensitive to noise and gives inaccurate load predictions when the dataset is only slightly noisy. In order to reduce the sensitivity of the ANN to noise, the training algorithm was modified by introducing noise to the training and validation datasets. The ANN was then trained, validated, and tested using noisy datasets with signal to noise ratios between 10 and 100 (Figure 11). It was found that when ANNs are trained using noisy datasets, they are very robust and work well even when the density measurements are noisy. The accuracy of the load prediction obviously decreases as the signal to noise ratio decreases (Figure 11). However, the reduction of accuracy with the increase of noise is gradual and simply reflects the lower quality of the measurement data. In short, training ANNs with noisy training datasets makes the load prediction algorithm robust against measurement noise. That is needed in order to use the proposed technique for prediction of the load from a density distribution that is measured, for example, by CT.

We have not yet investigated the worst-case scenario: when both noise and non-uniqueness are present. In order to assess the performance of ANNs in the worst-case scenario, ANNII was retrained using noisy data points. Although the performance of ANNII is not as good as in the noise-free case (Figure 12), the network still works reasonably well in prediction of the loading parameters. The distribution of the identification error for the test dataset was close to a normal distribution (Figure 12d). The mean absolute identification error was always less than 9% of the size of the identification interval (Table 3). Since the density distributions used for the training of the ANN contained a random component amounting to about 5% of the signal ($\lambda=20$), the performance of the ANN in identification of the loading parameters cannot be expected to be very accurate. A sample set of loading parameters was selected to assess how close is the density distribution that is obtained using the identified set of loading parameters to the density distribution that is obtained using actual loading parameters. The sample set of loading parameters was selected in the following way. For all the entries of the test dataset, the average of the identification errors of the four parameters ($F1-F4$) was calculated. Within all the entries, the one whose average of identification error was the least distant from the mean absolute identification error of the ANN (Table 3, column 3) was picked as the sample. For this sample set, the forward bone adaptation model was simulated using both *actual* ($F1=4.8$, $F2=4.6$, $F3=2.4$, $F4=3.0$) and identified ($F1=4.7$, $F2=4.5$, $F3=2.6$, $F4=2.9$) *loading* parameters. The density distributions obtained using the actual and identified loading parameters were found to be very close to each other (Figure 12e).

It should be noted that the problem investigated in this study was specifically designed to be as close to the worst-case scenario as possible. First, the number of elements used for solving the forward problem was limited to 25 elements. Distinguishing between different density distributions is more difficult when a limited number of elements are used in the solution of the forward problem, because then ANN has access to only a limited amount of information about the morphology of the bone. Even though distinguishing between two slightly different load cases is easier when the number of elements is larger, the computational cost of load prediction is also significantly higher. It is important to mention that the proposed load prediction technique was successfully applied to the case of $10\times10$ and $20\times20$ elements. Moreover, the occurrence of non-unique solutions is more likely when the number of elements is limited. In addition to using a limited number of elements, the chosen geometry and also the chosen loading were very symmetric. The symmetry of the geometry and loading makes it more likely to have non-unique solutions in the solution of the forward problem. The actual shapes of most bones (e.g. femur) are much less symmetric. The loading patterns are also not symmetric at all. Therefore, it is expected that the proposed load prediction technique would work even better in most other applications.

## 4. Implications for biomechanical research

The approach proposed in this study allows us to predict the loads that are experienced by bone tissue. Even though we only considered trabecular bone in the current study, the proposed methodology and algorithms can be also used for predicting the loads at the organ level (e.g. proximal femur) or even in a multi-scale scheme [21]. When the density distribution is measured at the organ level, the macro-scale averaged density distributions can be used to predict the musculoskeletal loads such as joint reaction and muscle forces. In an even more sophisticated approach, a multi-scale model [21] can be built to connect the different scales and to predict the musculoskeletal loads from the microstructure of bone using an approach similar to the one proposed in the current study.

The ability to predict the musculoskeletal loads from the density distribution of bones creates an important opportunity in biomechanical research. Until now, there has not been any non-invasive way for *in-vivo* measurement of musculoskeletal loads. Currently, the only feasible way for measuring musculoskeletal forces is implanting an instrumented prosthesis [37-39] in a patient's body.

In absence of readily available experimental techniques for measurement of musculoskeletal loads, one has to resort to modeling approaches for prediction of musculoskeletal loads. The modeling approaches that can be used for this purpose include large-scale musculoskeletal models [40-42] or simpler models such as mass-spring-damper models of the human body during physical activities [43-45]. However, the predictions of most such models are not validated against experimental measurements of musculoskeletal loads [46]. The computational approach proposed in the current study can be useful for validation of the predictions of large-scale musculoskeletal models. For such a validation study, one can use the musculoskeletal model to predict the musculoskeletal loads of a certain number of subjects. CT images will be also collected for the same subjects. The approach proposed in this study (perhaps applied at the organ or multi-scale level) can be used to predict the loads that have resulted in the measured density distribution. The musculoskeletal loads predicted with the musculoskeletal model can be then compared with the musculoskeletal loads identified through solving the inverse tissue adaptation problem.

The performance of the ANNs in prediction of loads from the density distribution of bone is not expected to be dependent on the parameters of the bone tissue adaptation model expect for the parameters that influence the distinguishability of different density distributions. There are two special sets of parameters that influence the distinguishability of density distribution. The first set of parameters includes the parameters that determine the extent of the diffusion of the bone remodeling stimulus signal. For example, when the parameter $D$ increases, the domain of influence of sensors (osteocytes) in the bone tissue adaptation model increases, resulting in a more diffused predicted density distribution (Figure 2). The performance of the ANN may suffer from a more diffused density distribution, because it would be more difficult for the ANN to distinguish between two density distributions that result from two different (but possibly close) sets of loading parameters. The second set of parameters includes the parameters that limit the extent of tissue adaptation. For example, in some versions of the Huiskes' tissue adaptation model [47], the bone tissue adaptation is considered to not happen until the remodeling stimulus signal reaches a certain threshold that defines the so-called "lazy zone" or "dead zone". When a lazy zone is implemented in the bone tissue adaptation model (not the case in the current study), two sets of loading parameters that are only slightly different result in the same distribution. The ANN cannot therefore distinguish between the two different sets of loading parameters. The same holds true when a certain maximum and/or a minimum density value are used in the bone tissue adaptation model.

The computational approach proposed in this study has the potential to be used at the trabecular scale, organ scale, and in multi-scale models. However, in this study the proposed methodology was only applied for prediction of the load from two-dimensional trabecular level density distribution. For exploiting the full potential of the method, it is important to apply it to three-dimensional problems at all organ/tissue scales including multi-scale models. Moreover, actual measured CT images must be used in the future to identify the musculoskeletal loads at the organ level. The predicted loads should be compared with the loads measured *in-vivo* using instrumented prosthesis or predicted by large-scale musculoskeletal models. Finally, the effects of different parameters such as age, gender, the anatomical location of the bone, and biological/systemic factors need to be studied in detail. Even though the proposed computational approach can be still used in these three suggested cases, it is important to actually apply the computational approach in such cases to demonstrate and exploit the full potential of the approach. Moreover, there may be a need for some extensions to the proposed approach when it is applied to certain biomechanical problems. For example, the finite element model may need to be adapted to account for certain biological aspects including the fact that bone evolves differently in different anatomical locations. In the specific case of different evolution of bone in different anatomical locations, one may, for example, need to make the value of parameter $k$ (Equation (3) and (7)) location-dependent. Furthermore, more sophisticated bone tissue adaptation models may be needed for capturing such effects as fatigue damage accumulation or interstitial fluid flow. Nevertheless, the overall structure of the proposed computational approach will remain unchanged. There will be still an ANN that needs to be trained using the solutions of the forward tissue adaptation problem. The only things that will change are the details of the formulation of the forward tissue adaptation model.

**4. Conclusions**

The main conclusion of this study is that artificial neural networks can be used for prediction of loading parameters that have resulted in a given density distribution. The involved inverse problem (going from the density distribution to loading parameters) is an ill-posed problem, because its solution is not unique. Two algorithms were proposed in this study to cope with the non-uniqueness of the solution of the inverse problem. Two ANNs were created using the proposed algorithms for preparation of the training datasets. While the first ANN specialized in prediction of unique loading parameters, the second ANN was specially designed for prediction of non-unique loading parameters. It was shown that both ANNs perform well in prediction of loading parameters from density distribution.

The effects of noise on the performance of the ANN were also investigated. It was shown that the ANN would give very inaccurate results when slightest degree of noise was introduced to the density distribution. It was therefore suggested that noise should be introduced to the training data points and the ANN should be trained using noisy data points. It was observed that the proposed modification in the training algorithm makes the ANN very robust against noisy density measurements. The effects of presence of both non-uniqueness and noise on the performance of the ANNs were also investigated. It was shown that the ANN works reasonably well even when both difficulties are present.

**Figure captions**

**Figure 1.** The two types of geometry, boundary conditions, and loading used in this study. The problem presented in subfigures (a) is similar to the one studied in [32] and the problem presented in subfigure (b) is similar to the one considered in [28].

**Figure 2.** The density distribution calculated using non-normalized stimulus signal (a) and normalized stimulus signal (b). The non-normalized and normalized stimulus signals are respectively calculated using Equations (3) and (7).

**Figure 3.** A schematic drawing of the architecture of a typical feedforward artificial neural network (a) and one single hidden neuron (b).

**Figure 4.** Identified vs. actual loading parameters for the training (a), validation (b), and test (c) datasets. The training curve of the ANN (d) when a total of 100 training samples (including training, validation, and test datasets) are used.

**Figure 5.** Identified vs. actual loading parameters for the training (a), validation (b), and test (c) datasets. The training curve of the ANN (d) when a total of 1000 training samples (including training, validation, and test datasets) are used. The absolute identification errors for the test dataset are presented in a histogram (e).

**Figure 6.** Identified vs. actual loading parameters for the training (a), validation (b), and test (c) datasets when non-unique solutions are present.

**Figure 7.** The percentage of deleted training samples vs. the applied similarity threshold for a dataset consisting of 65000 samples.

**Figure 8.** Identified vs. actual loading parameters for the training (a and d), validation (b and e), and test (c and f) datasets when non-unique solutions are filtered out using the first algorithm.

**Figure 9.** Identified vs. actual loading parameters for the training (a), validation (b), and test (c) datasets when only non-unique solutions are used for the training of the network (ANNII).

**Figure 10.** Identified vs. actual loading parameters when an ANN trained using noise-free training data is used for prediction of density distributions with noise ($\lambda = 100$).

**Figure 11.** Identified vs. actual loading parameters for the training (a), validation (b), and test (c) datasets when neural networks are trained using noisy training datasets with increasing levels of noise.

**Figure 12.** Identified vs. actual loading parameters for the training (a), validation (b), and test (c) datasets when both non-uniqueness and noise ($\lambda = 20$) are present in the training dataset of ANNII. The absolute identification errors for the test dataset are presented in a histogram (d). A comparison of the density distribution obtained from the actual loading parameters with the density distribution obtained from the identified loading parameters for one sample set of loading parameters (e).

**Table captions**

**Table 1.** The parameters used for the simulations presented in Figure 2. The first four parameters (first four columns) are used also in most other simulations of this study.

**Table 2.** Statistical analysis of the vector of identification errors (only the test dataset is considered). The results are related to the ANN whose results are presented in Figure 5.

**Table 3.** Statistical analysis of the vector of identification errors (only the test dataset is considered). The results are related to the ANNII whose results are presented in Figure 12.

**Table 1**

| $k$ [J / g] | $C$ [$MPa / (gcm^{-3})^2$] | $\gamma$ | Sensors | $F$ [$N / mm^2$] | $T$ [$N / mm^2$] | Intervals | Mass [$g$] |
|---|---|---|---|---|---|---|---|
| 0.25 | 100 | 2 | 40000 | 10 | 3.5 | 200 | 411 |

**Table 2**

|  | $\mu_{err}$ | $\sigma_{err}$ | $\mu_{abs} / (F_{max} - F_{min}) \times 100$ |
|---|---|---|---|
| $F1$ | -2.932e-04 | 0.0035 | 0.1356% |
| $F2$ | 0.0012 | 0.0052 | 0.1826% |
| $F3$ | -0.0012 | 0.0051 | 0.1766% |
| $F4$ | -1.512e-04 | 0.0039 | 0.1600% |

Table symbols: $\mu_{err}$: mean identification error, $\sigma_{err}$: standard deviation of identification error, $\mu_{abs} / (F_{max} - F_{min}) \times 100$: mean absolute identification error divided by the size of the identification interval.

**Table 3**

|  | $\mu_{err}$ | $\sigma_{err}$ | $\mu_{abs} / (F_{max} - F_{min}) \times 100$ |
|---|---|---|---|
| $F1$ | -0.0164 | 0.1663 | 4.23% |
| $F2$ | 0.0496 | 0.3697 | 8.17% |
| $F3$ | 0.0216 | 0.2839 | 6.85% |
| $F4$ | 0.0035 | 0.1810 | 4.37% |

Table symbols: $\mu_{err}$: mean identification error, $\sigma_{err}$: standard deviation of identification error, $\mu_{abs} / (F_{max} - F_{min}) \times 100$: mean absolute identification error divided by the size of the identification interval.

**Figure 1**

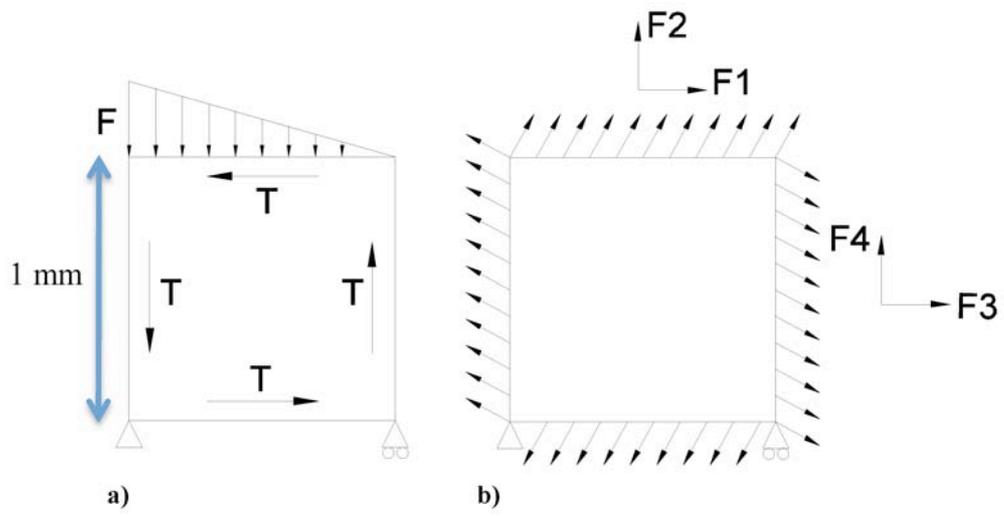

**Figure 2**

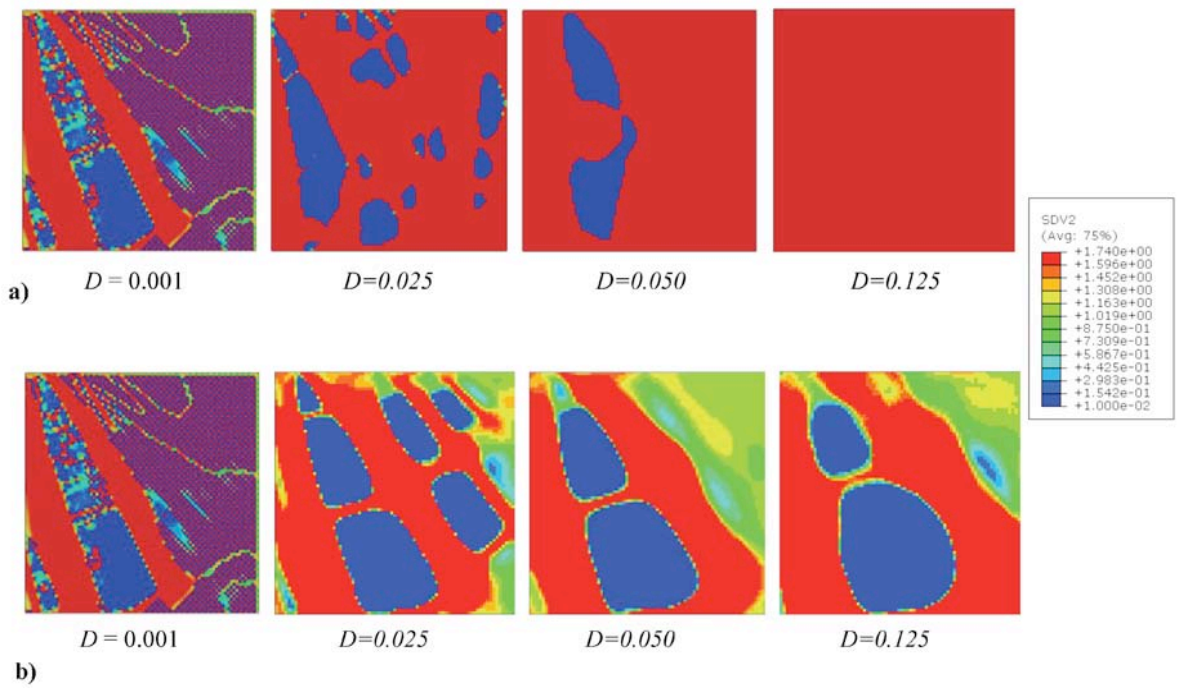

**Figure 3**

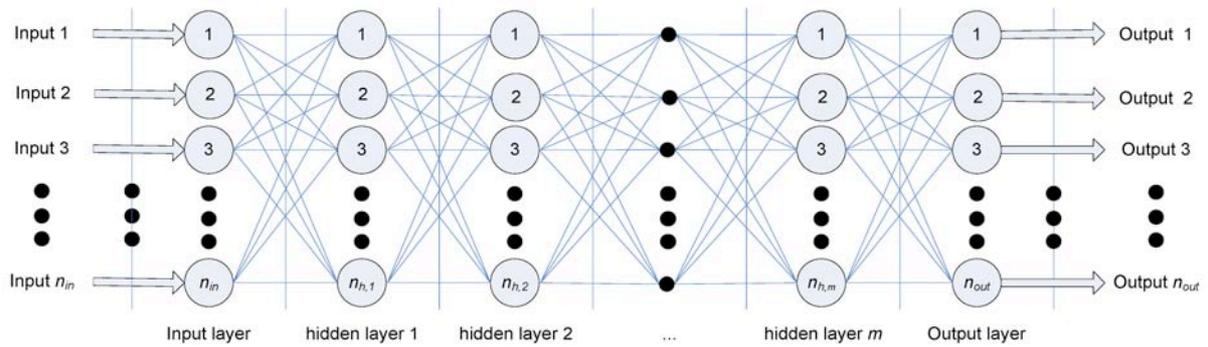

a)

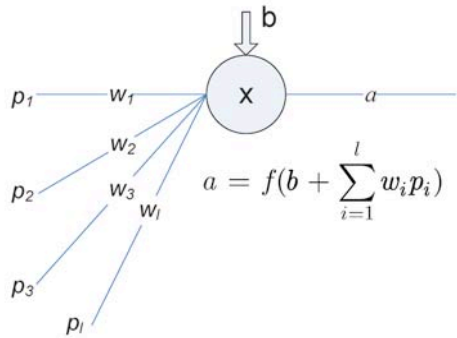

b)

**Figure 4**

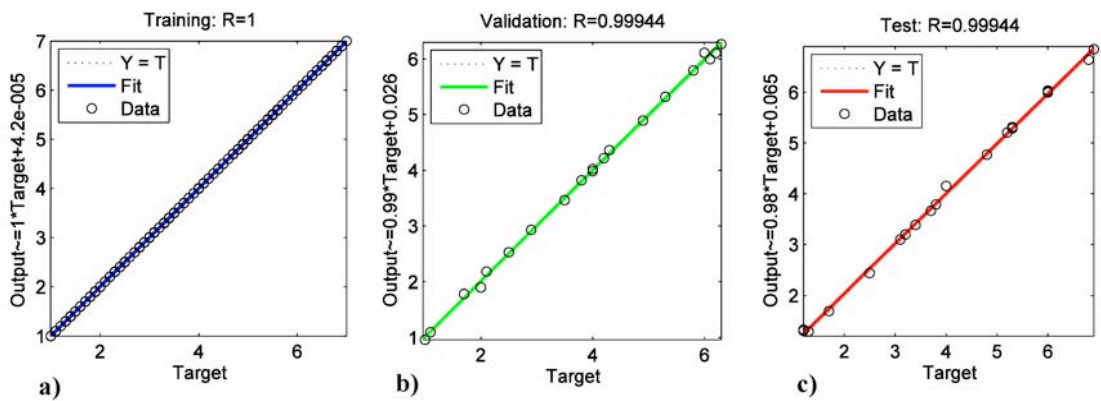

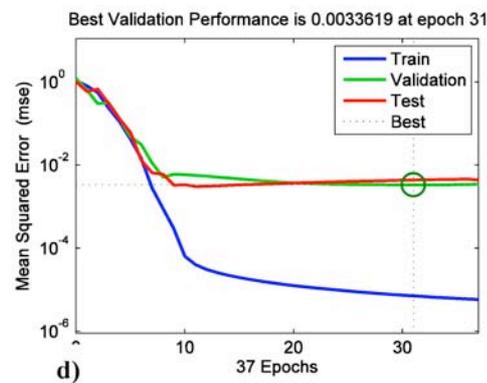

**Figure 5**

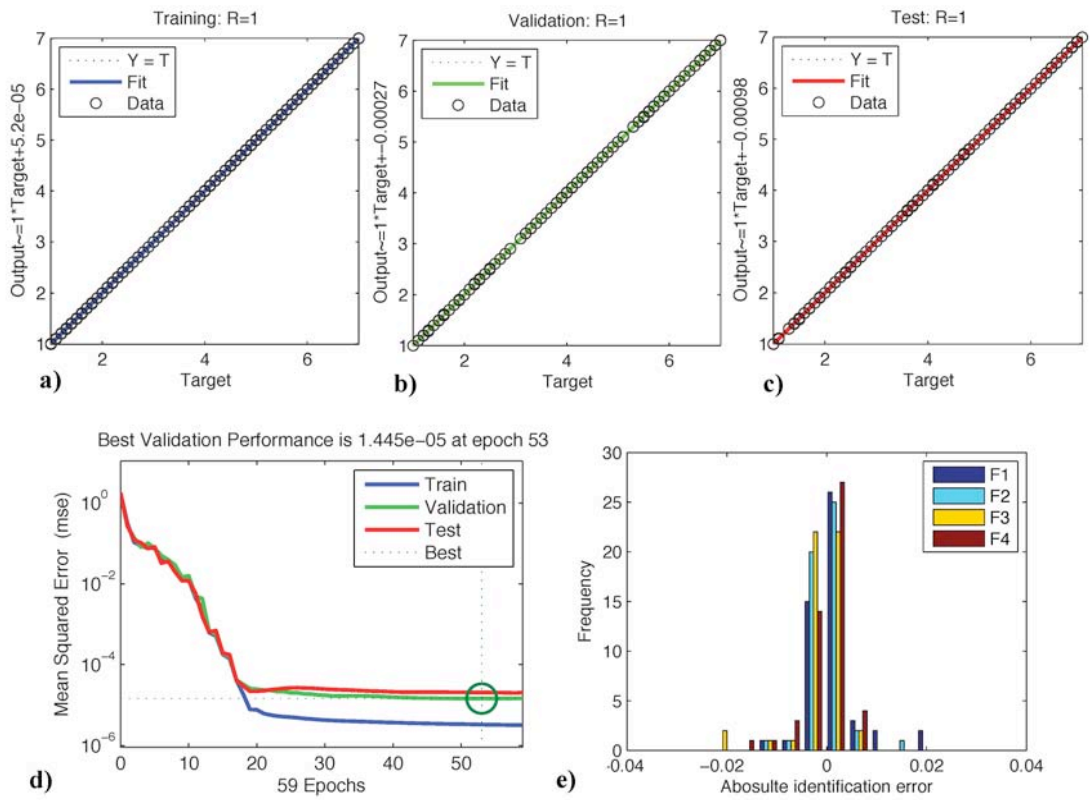

**Figure 6**

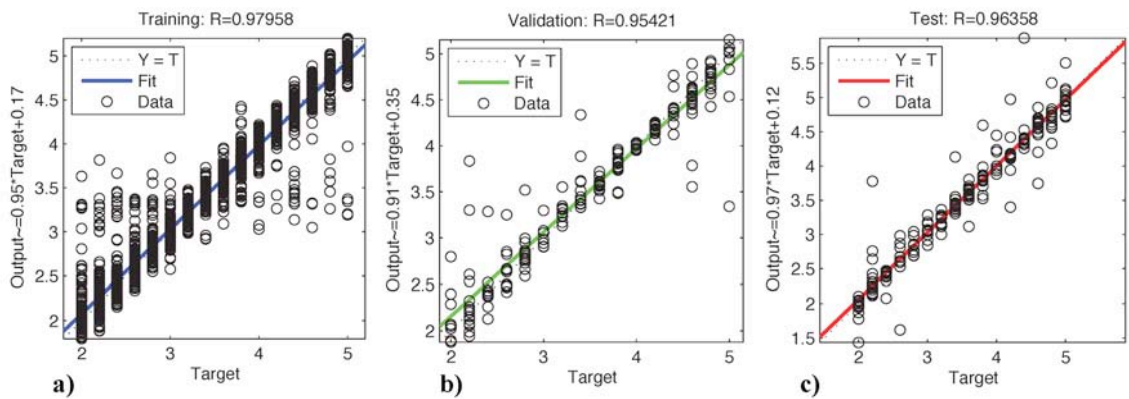

| No. samples | No. neurons |
|---|---|
| 1000 | 20 |

**Figure 7**

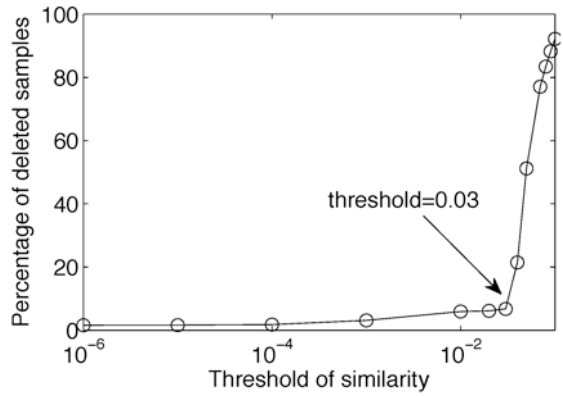

**Figure 8**

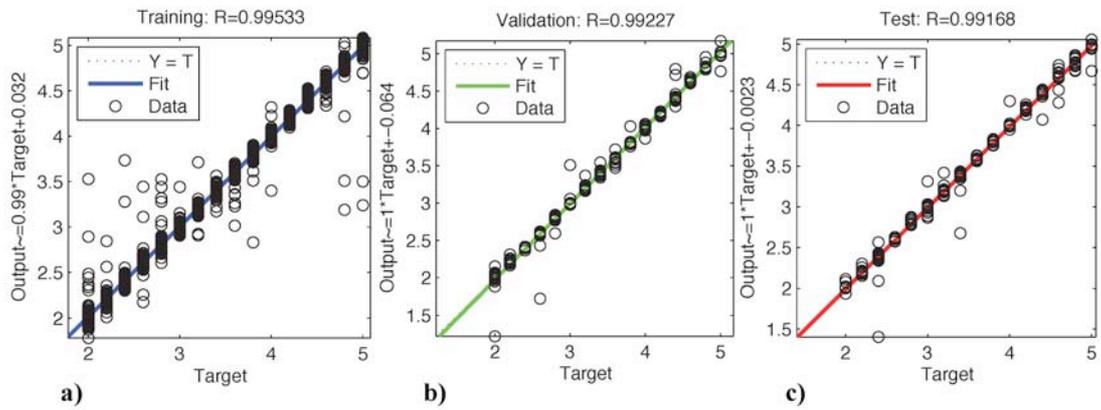

| No. samples | No. neurons | $l_{th}$ |
|---|---|---|
| 1000 | 20 | 0.03 |

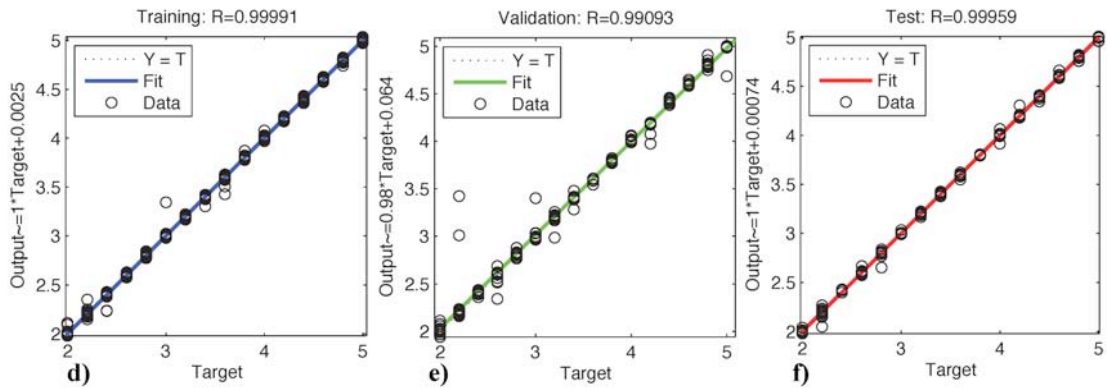

| No. samples | No. neurons | $l_{th}$ |
|---|---|---|
| 1000 | 60 | 0.07 |

**Figure 9**

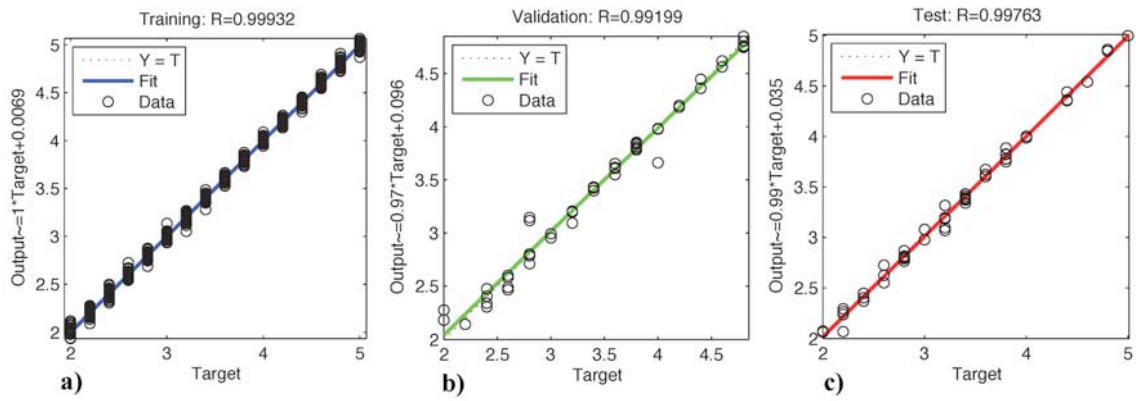

| No. samples | No. neurons | $l_{th}$ | $L_{th}$ |
|---|---|---|---|
| 267 | 10 | 0.1 | 0.4 |

**Figure 10**

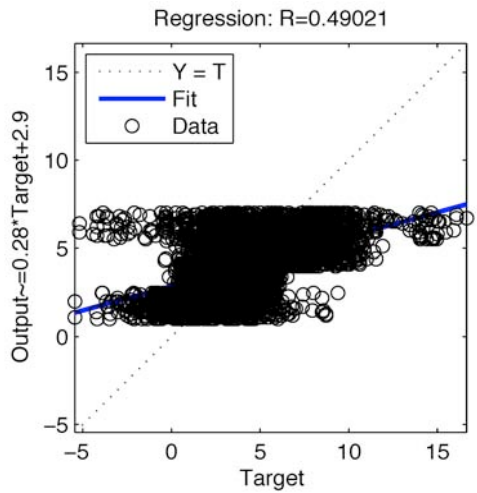

**Figure 11**

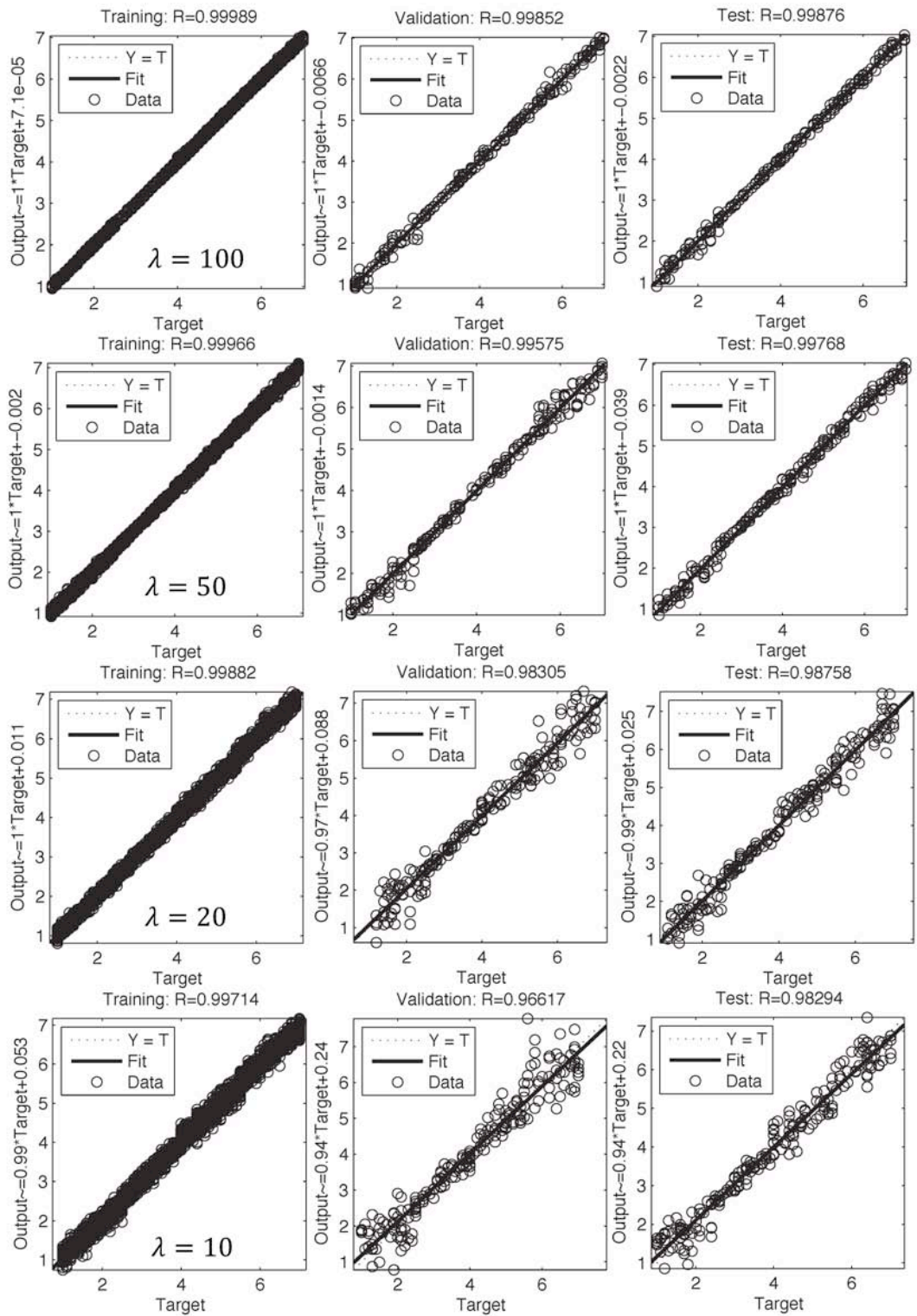

**Figure 12**

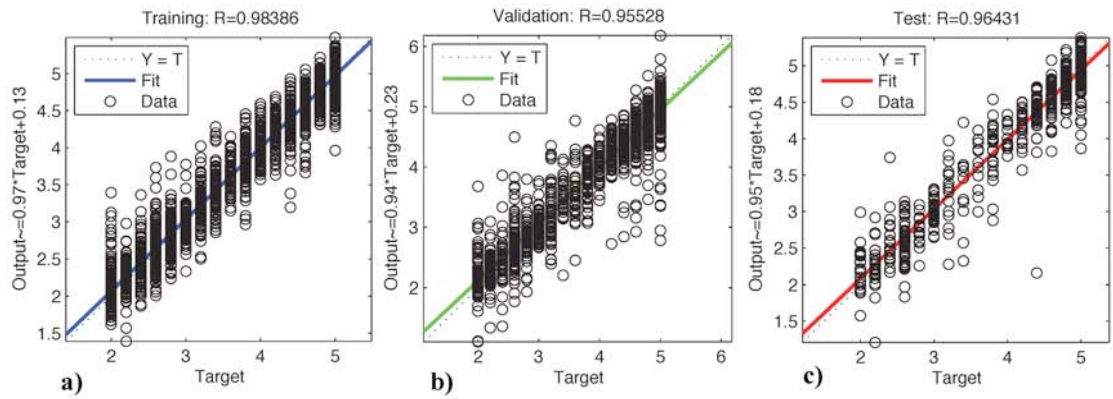

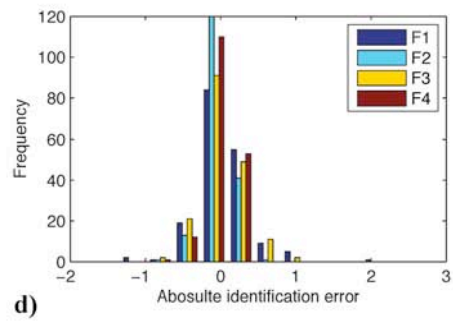

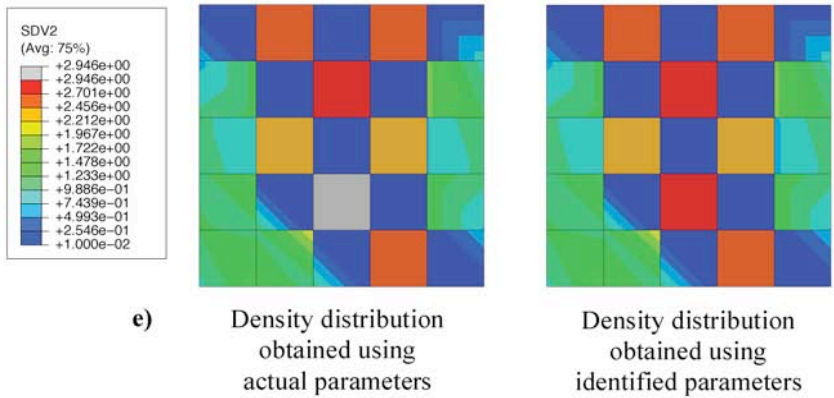